\documentclass[10pt, sort&compress, a4paper]{article}

\usepackage{amssymb}
\usepackage{mathtools}
\usepackage{mathrsfs}
\usepackage{enumitem}
\usepackage[square,sort,comma,numbers]{natbib}
\usepackage{gensymb}
\usepackage{url}
\usepackage{caption}
\begin{document}
\title{Uncertainties in ROC (Receiver Operating Characteristic) Curves Derived from Counting Data}
\author{M.P. Fewell \cr ARC Centre of Excellence for Dark-Matter Particle Physics \cr and \cr Department of Physics, The University of Adelaide, \cr Adelaide SA 5005, Australia}
\date{June 2024}
\maketitle
\begin{abstract} \noindent
The ROC (receiver operating characteristic) curve is a widely used device for assessing decision-making systems. It seems surprising, in view of its history dating back to World War Two, that the assignment of uncertainties to a ROC curve is apparently not settled. This note returns to the question, focusing on the application of ROC curves to the analysis of data from counting experiments and taking a practical operational approach to the concept of uncertainty. 
\end{abstract}

\section{Introduction}
\label{sec:Introduction} \noindent
The ROC (receiver operating characteristic) curve is a tool for assessing and tuning data-analysis procedures, in particle physics and much more widely; for it is a fundamental construct of signal-detection theory, applying to decision-making under uncertainty, whenever one seeks to balance the false-alarm rate against the rate of missed detections. As commonly plotted, a ROC curve shows the probability of a detection\textemdash the complement of the probability of a missed detection\textemdash as a function of the probability of a false alarm (where a detection is declared although the looked-for object or occurrence is in fact absent).

ROC curves are constructed by comparing two classes of events: those in which a desired signal is known to be present, for example by construction in a simulation of the process producing the events, and those where it is known to be absent. For each class, one defines the \emph{efficiency} as the fraction of events in which a detection is declared. A good detection process has high efficiency for events containing signals\textemdash a high detection probability\textemdash and simultaneously low efficiency for non-signal events\textemdash a low false-alarm (or false-positive) rate. The two efficiencies are the coordinates of a point on the ROC curve. The full curve is the locus of such points as some process parameter that affects the efficiencies is varied.

This note deals with ROC curves constructed from counting data, taking an example from experimental elementary-particle physics. The process parameter affecting the efficiencies is a threshold, with the same threshold being applied to signal and non-signal data. Sections \ref{sec:Effmeth} and \ref{sec:ROCconstr} develop an example.

The assignment of uncertainties to ROC curves is important, for often one wishes to compare ROC curves.  It implies assigning uncertainties to the efficiencies that are the coordinates of points on a ROC curve. This seems to be an open question (e.g. \citep{He09}), perhaps surprisingly. The present note argues that a previous treatments of the question in the context of counting data, such as those by Ullrich and Xu \citep{Ul07}\textemdash supposed in \citep{He09} to be correct\textemdash or by Casadei \citep{Ca12}, deal with a different type of efficiency from that used in ROC-curve construction, a difference which appears to impact the statistical behaviour.

The calculation of the type of efficiency needed for ROC-curve construction from counting data is described in \S\ref{sec:Effmeth}. Section \ref{sec:Effunc} computes the uncertainty in such an efficiency using an operationally inspired approach that aims to mirror the calculational method closely, having regard to the practical application of the resulting uncertainties. The approach makes no assumption about the statistical behaviour of efficiency\textemdash such as, for example, that it is binomially distributed\textemdash nor does it require that a Bayesian prior be postulated.  The approach does depend on the statistical independence of two particular quantities; this is argued in \S\ref{sec:IndepPF}, which begins with a conceptualisation of the nature, source and uses of uncertainty in the present context.

Implications for interpreting the resulting uncertainties are set out in \S\ref{sec:ROC}.  Section \ref{sec:UlXu} considers specific criticisms by Ullrich and Xu \citep{Ul07}, arguing that they do not apply to ``efficiency`` calculated as described in \S\ref{sec:Effmeth}, and concluding in \S\ref{sec:Aside} with the important observation that the points near the ends of ROC curves are of negligible interest when comparing ROC curves.

\section{Efficiency as Used in ROC Curves}
\label{sec:Eff}
\subsection{Efficiency Calculation}
\label{sec:Effmeth} 
\begin{figure}
    \centering
    \includegraphics[width=0.9\textwidth]{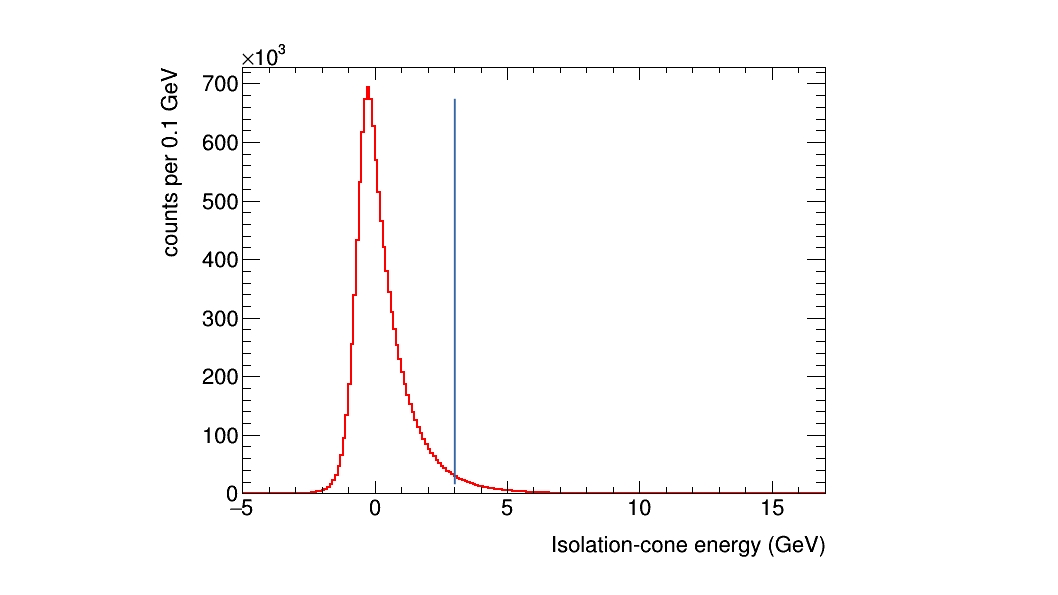}
    \caption{Histogram of energy in an isolation cone around a photon candidate in the ATLAS detector. Values less than zero reflect the effect of detector energy resolution on the computation isolation energy. The blue line shows an example of a cut used to compute efficiency, here at 3.0 GeV. In this application, events with energy less than the cut are declared to be isolated; that is, such events pass the isolation test.}
    \label{fig:one_histo}
\end{figure}

\noindent
Figure~\ref{fig:one_histo} shows a histogram of isolation-cone energy\footnote{Several definitions of isolation-cone energy are used in ATLAS. The differences are unimportant for present purposes: all definitions have the same underlying statistical properties. It is also unimportant statistically that the cut is often applied to the ratio of isolation-cone energy to transverse photon momentum, rather than to isolation-cone alone. The histogram in Fig.~\ref{fig:one_histo} shows results of Monte-Carlo simulation of the detection process, rather than data from actual proton\textendash proton collisions. This can have statistical implications, but is unimportant for the illustrative purpose to which it is applied here.} for cones of a particular size surrounding photon candidates in the ATLAS detector at CERN. To achieve isolation, one stipulates that the energy in the cone be less than a certain value, called the \emph{threshold} or \emph{cut}; an example is marked in Fig. \ref{fig:one_histo}. Events falling below the cut are declared to pass the isolation test. The efficiency $\varepsilon$ of the test is simply
\begin{equation} \label{effNorm}
    \varepsilon = \frac{P}{N},
    \end{equation} 
\noindent where $P$ is the number of events passing the test and $N$ is the total number of events in the histogram.

Obviously $\varepsilon$ varies with the location of the cut, behaviour that is foundational to the ROC-curve concept. This is developed in \textsection \ref{sec:ROC}, but first we consider the uncertainty in $\varepsilon$.

\subsection{Uncertainty in Efficiency}
\label{sec:Effunc} \noindent
For uncertainty determination, eq. \eqref{effNorm} has the difficulty that $P$ and $N$ are correlated. So, let us rewrite $\varepsilon$ as
\begin{equation} \label{effMod}
    \varepsilon = \frac{P}{P+F}\,,
    \end{equation}       
\noindent where $F$ is the number of events failing the test. Section \ref{sec:IndepPF} argues that $P$ and $F$ are statistically independent. Given that, the standard rule for propagating uncertainties in quadrature can be applied: the uncertainty $\Delta\varepsilon$ in the efficiency is given by
\begin{equation} \label{uncProp}
    (\Delta\varepsilon)^2 = \left (\frac{\partial\varepsilon}{\partial P}\right )^2(\Delta P)^2 + \left (\frac{\partial\varepsilon}{\partial F}\right )^2(\Delta F)^2.
    \end{equation}
\noindent Carrying out the differentiations gives
\begin{equation} \label{effWtunc}
    (\Delta \varepsilon)^2 = \frac{F^2\,(\Delta P)^2 + P^2\,(\Delta F)^2}{(P+F)^4}\,.
    \end{equation}
\noindent If the events are equally weighted,\footnote{This need not be the case for events generated by Monte-Carlo simulation \citep[\S7.2.2]{Bu21}.} Poisson-distributed and sufficiently numerous for the Gaussian approximation to apply, then
\begin{equation} \label{varUncs}
(\Delta P)^2 = P, \qquad (\Delta F)^2 = F,
\end{equation}
\noindent which, with eq. \eqref{effMod}, reduces eq. \eqref{effWtunc} to
\begin{equation} \label{unc}
    (\Delta\varepsilon)^2 = \frac{\varepsilon (1-\varepsilon)}{N}.
    \end{equation}
    
Equation \eqref{unc} implies that $\Delta\varepsilon$ goes to zero as $\varepsilon \rightarrow 0$ and as  $\varepsilon \rightarrow 1$, behaviour that has been criticised as ``absurd'' and that ``violates our reasonable expectation'' \citep[\textsection 2.2]{Ul07}. It is argued in \textsection \ref{sec:UlXu} that this is not so: the behaviour is reasonable and ought to be expected for the type  of efficiency under discussion here. 

\subsection{Statistical Independence of $P$ and $F$}
\label{sec:IndepPF} \noindent
From an operational perspective, the purpose of uncertainty is to indicate the extent to which the result may change if the whole analysis is repeated: for a histogram built from experimental data, if the experiment is repeated; for one built from simulation, if the Monte-Carlo simulation is repeated from scratch. An uncertainty should reflect the fluctuation in results among many such repetitions. Statistical correlation of two quantities then refers to their relationship from repetition to repetition: whether they tend to fluctuate together, either in synchrony or contrary-wise (positive or negative correlation respectively), or not (the case of independence).

For the numbers of events passing and failing the test\textemdash $P$ and $F$ respectively (\textsection \ref{sec:Effunc})\textemdash the argument for their statistical independence proceeds in two steps: (i) showing that bins in a histogram like that in Fig. \ref{fig:one_histo} are statistically independent of each other and (ii) establishing conditions for the statistical independence of different sums over histogram bins.

\renewcommand{\labelenumi}{\textbf{(\roman{enumi})}}
\begin{enumerate}
\setlength{\parindent}{1.5em} \setlength{\parskip}{0pt plus1pt}
\item \textbf{Statistical independence of individual histogram bins.} Choose a bin in the histogram\textemdash say at 2.0~GeV in Fig.~\ref{fig:one_histo}\textemdash and, from many repetitions of the analysis that produces this histogram, choose those in which the contents of this bin exceeds the mean value over all repetitions. One asks: does the value in another bin\textemdash at 5.0~GeV, say\textemdash also exceed the mean for that bin (or fall below that mean) in the chosen histograms more often than not? If so, there is correlation between the two bins; if not, not; i.e. they are statistically independent. Now each count in any bin in Fig.~\ref{fig:one_histo} comes from a separate event: from a collision between two protons in the case of data, or from a simulation of such a collision in the other case. There is no memory from collision to collision; there is nothing to cause a correlation. This applies between any two bins in the histogram, establishing their statistical independence. 

Ullrich and Xu \citep{Ul07} seem implicitly to assume the statistical independence of individual histogram bins in the second-last sentence of their \textsection 1, so the argument of the previous paragraph would appear to be uncontroversial.

\item \textbf{Condition for statistical independence of different sums over histogram bins.} Since any two bins of a histogram are statistically independent, so will be two sums over histogram bins if the ranges of summation do not overlap. This must be so, for such a summation is no more than re-binning the histogram, and individual bins in the re-binned histogram are just as statistically independent of each other as are the bins of the original histogram.

If there is overlap, then bins in the region of overlap contribute to both sums, thereby providing a mechanism for correlation.  From the perspective of re-binning, one never includes an original bin in more than one new bin; for to do so would be double-counting.
\end{enumerate}
By construction, the sums $P$ and $F$ in \textsection \ref{sec:Effunc} have summation ranges that do not overlap. Hence they are statistically independent. On the other hand, the contrary applies to the quantities $P$ and $N$ in eq. \eqref{effNorm}: the summation range for $P$ is contained within that for $N$.

\section{ROC Curves} \label{sec:ROC} 
\subsection{Construction and Uncertainties}
\label{sec:ROCconstr} 
\begin{figure}[b!]
    \centering
    \includegraphics[width=0.9\textwidth]{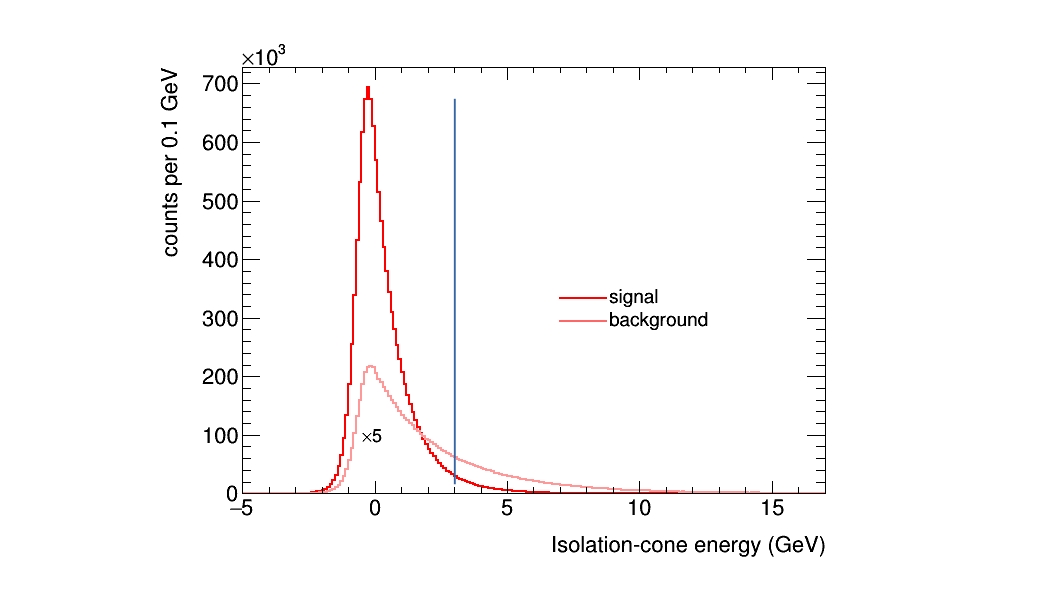}
    \caption{Like Fig. \ref{fig:one_histo}, but including "background" events known not to contain a signal photon, but which for various reasons have been misclassified as though they do. (The signal events are the same as in Fig. \ref{fig:one_histo}).}
    \label{fig:two_histos}
\end{figure}

\begin{figure}
    \centering
    \includegraphics[width=0.9\textwidth]{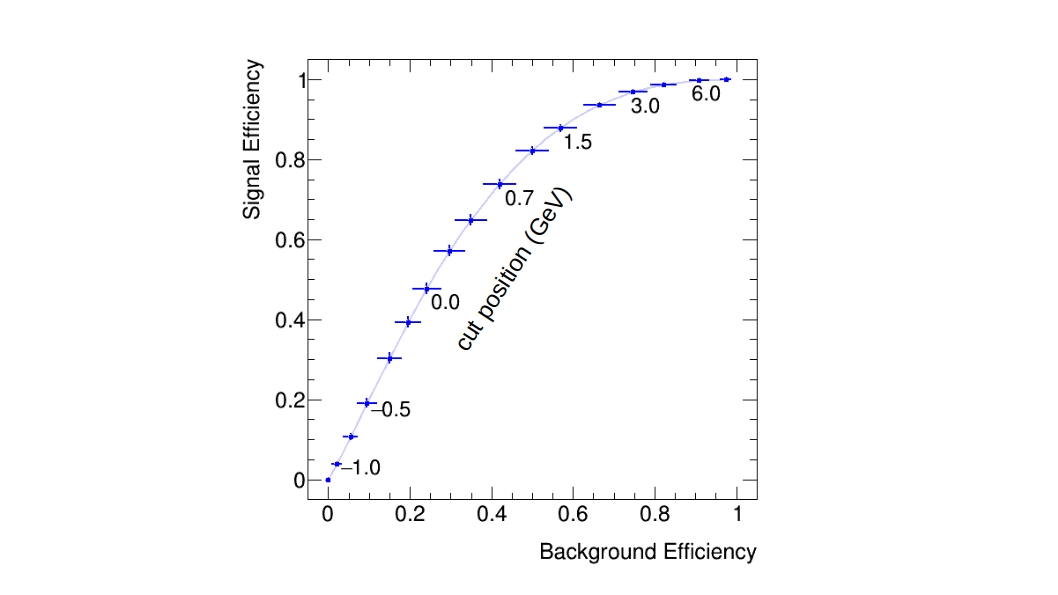}
    \caption{ROC curve constructed from Fig. \ref{fig:two_histos} by selecting a range of locations for the cut (shown at 3.0 GeV in Fig. \ref{fig:two_histos}) and computing the fractions of events, either signal or background, passing the test at each position. The markers show the cut positions chosen; the curve is a smooth interpolation. The error bars are 100 times the uncertainties given by eq. \eqref{unc}, with the $100\times$ scaling applied to make the error bars visible.}
    \label{fig:ROC1}
\end{figure}

\noindent
Figure \ref{fig:two_histos} shows an example of histograms from which a ROC curve can be constructed.  The construction consists in choosing locations of the cut\textemdash shown in Fig. \ref{fig:two_histos} at 3 GeV\textemdash over the full range of values and computing the fraction of the histogrammed events passing the test at each cut location. That is, the cut location parameterises the ROC curve. The result is Fig. \ref{fig:ROC1}, in which the points show the cut locations actually chosen, some of which are labelled, and the curve is a smooth interpolation. 

The error bars in Fig. \ref{fig:ROC1} are 100 times the uncertainties calculated according to eq. \eqref{unc}. The scaling factor of 100 was needed to make the error bars appreciably larger than the markers.
  
Concerning statistical independence: the uncertainties in the x and y directions on any given marker are independent, but there is correlation from marker to marker.  That is, the x-direction uncertainties displayed in Fig. \ref{fig:ROC1} are partially correlated with each other, as are the y-direction uncertainties. This occurs because the coordinates of the markers in a given direction are obtained through different cuts on the same histogram.

\subsection{Metrics and their Uncertainties}
\label{sec:ROCunc} 
\begin{figure}[b!]
    \centering
    \includegraphics[width=0.9\textwidth]{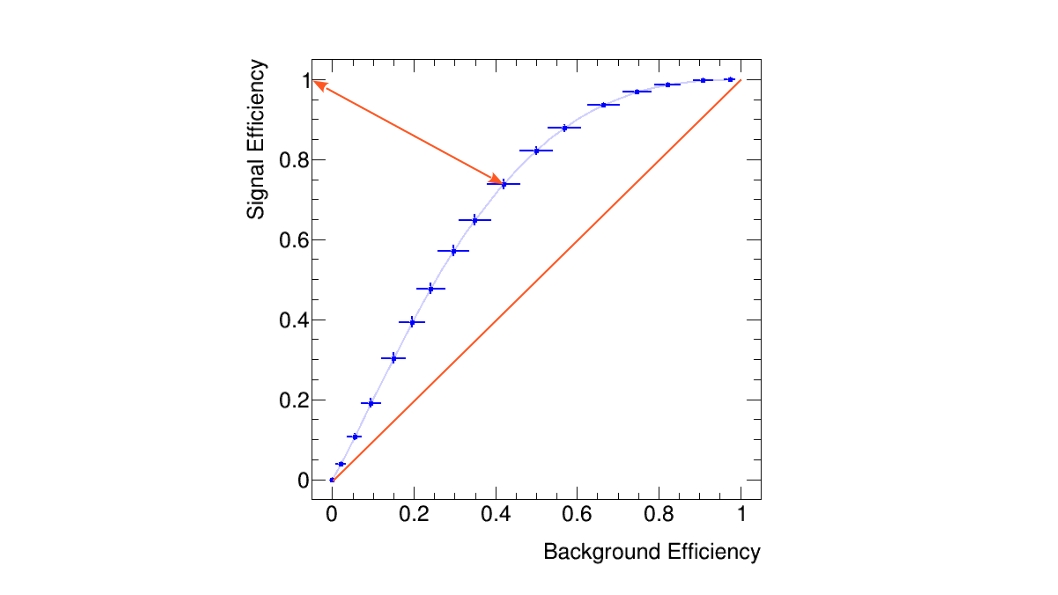}
    \caption{Like Fig. \ref{fig:ROC1}, but illustrating the ``distance $\mathscr{D}$ from perfection'' of a point on the ROC curve (arrow). The trailing diagonal is also shown.}
    \label{fig:ROC_metrics}
\end{figure}

\noindent
The \emph{area under the curve} is a commonly used metric for comparing ROC curves of different processes. It approaches unity as the system approaches perfect performance, which is signified by the ROC curve passing through the point (0, 1): zero background efficiency simultaneous with unit signal efficiency. It is a satisfactory metric, but the calculation of its uncertainty must take into account the correlation between uncertainties described in the last paragraph of \textsection \ref{sec:ROCconstr}. This awkwardness can be avoided by use of a different metric: minimum distance $\mathscr{D}_{min}$ from perfection. Figure \ref{fig:ROC_metrics} illustrates. The value of $\mathscr{D}$ for any point $(\varepsilon_\mathrm{bgd},\,\varepsilon_\mathrm{sig})$ on the ROC curve is simply

\begin{figure}
    \centering
    \includegraphics[width=0.9\textwidth]{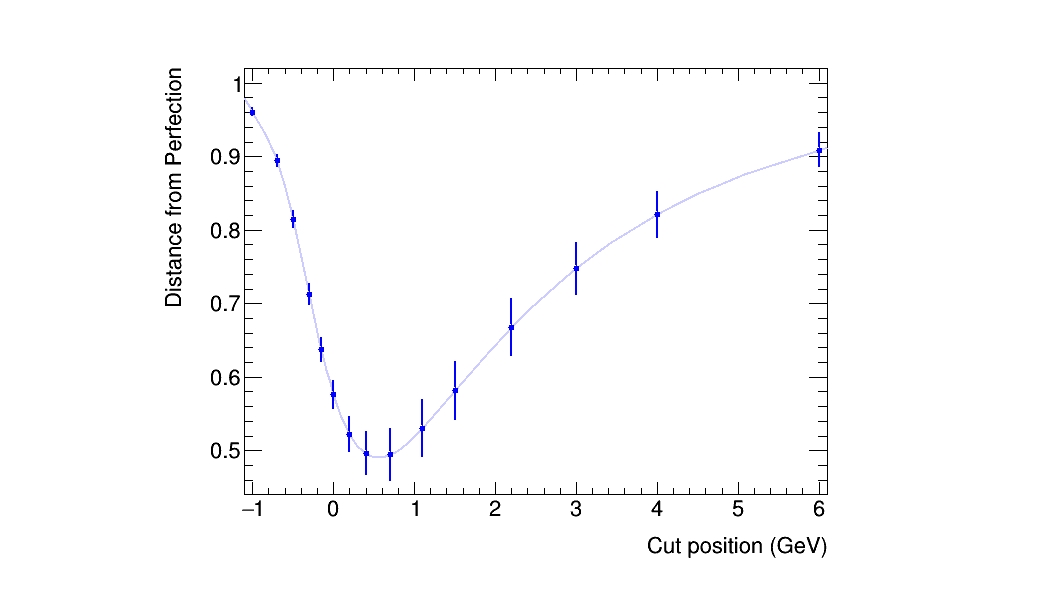}
    \caption{Variation of distance $\mathscr{D}$ from perfection along the ROC curve of Fig. \ref{fig:ROC_metrics}. The minimum value of this quantity is equivalent to ``area under the curve'' as a metric if quality of the detection process that produced the ROC curve.}
    \label{fig:bigDfig}
\end{figure}

\begin{equation} \label{bigD}
    \mathscr{D} = \sqrt{(\varepsilon_\mathrm{bgd})^2+(1-\varepsilon_\mathrm{sig})^2}\,.
    \end{equation}
\noindent In principle $\mathscr{D}$ could be as large as $\sqrt{2}$, but the ROC curves of practical processes rarely fall below the trailing diagonal, so in practice $\mathscr{D} \leq 1$ and $\mathscr{D}_\mathrm{min} \leq \sqrt{2}/2$. Figure \ref{fig:bigDfig} shows the variation of $\mathscr{D}$ along the curve in Figs \ref{fig:ROC1} and \ref{fig:ROC_metrics}. The minimum value is the metric of interest.

Because $\mathscr{D}$ involves just one point on the curve, deriving an expression for its uncertainty is straightforward.  The efficiencies $\varepsilon_\mathrm{bgd}$ and $\varepsilon_\mathrm{sig}$ are statistically independent, so the equivalent of eq. \eqref{uncProp} applies:
\begin{equation} \label{uncPropD}
    (\Delta\mathscr{D})^2 = \left (\frac{\partial\mathscr{D}}{\partial\varepsilon_\mathrm{bgd}}\right )^2(\Delta\varepsilon_\mathrm{bgd})^2 + \left (\frac{\partial\mathscr{D}}{\partial\varepsilon_\mathrm{sig}}\right )^2(\Delta\varepsilon_\mathrm{sig})^2,
    \end{equation}
\noindent which, when eq. \eqref{unc} applies, evaluates to
\begin{equation} \label{bigDunc}
    (\Delta \mathscr{D})^2 = \frac{\varepsilon_\mathrm{bgd}^3 (1-\varepsilon_\mathrm{bgd})}{\mathscr{D}^2 N_\mathrm{bgd}}+ \frac{(1-\varepsilon_\mathrm{sig})^3 \varepsilon_\mathrm{sig}}{\mathscr{D}^2 N_\mathrm{sig}}\,.
    \end{equation}

\section{Discussion \textemdash \ Response to the Criticisms of Ullrich and Xu}
\label{sec:UlXu} \noindent
Ullrich and Xu \citep{Ul07} derive eq. \eqref{unc} through use of the binomial distribution. As noted at the end of \S\ref{sec:Effunc}, they state that it is ``absurd'' to suppose the uncertainty to be zero when $\varepsilon$ is zero or unity, that the result is ``defective'' \citep[Abstract]{Ul07}, that it ``violates our reasonable expectation'' \citep[\S2.2]{Ul07}. They also claim that the total number $N \; (= P + F)$ of events ``is a fixed quantity and not subject to any fluctuation'' \citep[\S2.1]{Ul07}. The present section responds to these claims in turn, then proposes a resolution.

\subsection{Uncertainty at Zero and Unit Efficiency}
\label{sec:UncLim} \noindent
The discussion in \S\ref{sec:ROCconstr} indicates that, to obtain either zero or unit efficiency $\varepsilon$, the cut must be positioned at infinity, either positive or negative.\footnote{in principle. Real histograms, being discrete, do not extend to infinity in practice, but the underlying distributions do, in principle.} Consider repetitions of the analysis such as described in \S\ref{sec:IndepPF} with the cut at $+\infty$. Every repetition gives $\varepsilon = 1$ without exception; the uncertainty is indeed zero. This is appropriate in view of the meaning ascribed to ``uncertainty'' in the first paragraph of \S\ref{sec:IndepPF}.

A similar argument applies with the cut at $-\infty$: $\varepsilon = 0$ for every repetition, so the uncertainty is again zero.

One concludes not only that eq. \eqref{unc} is correct when $\varepsilon = 0$ or 1, but also that any analysis giving a nonzero uncertainty for either of these $\varepsilon$ values must be incorrect.

\subsection{Statistical Nature of Total Number $N$ of Events}
\label{sec:N} \noindent
Ullrich and Xu state that $N$ ``is a fixed quantity and not subject to any fluctuation. It's usually a well defined and known input quantity.'' \citep[\S2.1]{Ul07} Section \ref{sec:IndepPF} makes it clear that this cannot apply to the quantity $N$ as used herein; for $N=P+F$ and not only do $P$ and $F$ fluctuate from repetition to repetition, but also we have the result that they are statistically independent, so their sum must fluctuate too.

Fixed $N$ is a feature of the binomial distribution. That is, for the binomial distribution to apply in a given situation, not only must the result of an event be binary, but the number of events in a trial must be constant and specified. Repetition consists of repeated trials. The second condition does not apply to counting data used for ROC-curve construction.

These considerations hint that the analytical situation underlying the calculation of ROC curves is not the same as that considered by Ullrich and Xu, at least as regards statistical behaviour. This suggestion is developed in the next subsection.

\subsection{Synthesis}
\label{sec:Syn} \noindent
Referring to the calculation of efficiency, Ullrich and Xu write \citep[\S1]{Ul07}
\begin{quote}
\textellipsis this procedure can be simplified to the comparison of two histograms. In histogram A, one plots the distribution of the quantity of interest for all the data of the sample; in histogram B one plots the distribution of the same quantity, but only for those satisfying the selection criteria, \emph{i.e.}, those data that pass the cuts. Intuition leads one to expect that the best estimate for the (unknown true) efficiency for each bin is just $k_i/n_i$ where $k_i$ is the number of entries in bin $i$ of histogram B and $n_i$ is the number of entries in bin $i$ of histogram A.
\end{quote} \noindent
Clearly, the authors envisage the selection process\textemdash the definition of the cuts\textemdash \ as being separate from the quantity being histogrammed: the selection is determined by some other quantity or quantities. This is different from the selection process involved in the construction of a ROC curve. The difference seems too slight to be material, but it is a second indication of a divergence between analytical situations.

The efficiency treated by Casadei \citep{Ca12} has the same feature, as he states several times, for example in the first sentence of his \S4: ``\textellipsis\, the events had been selected by an independent process whose efficiency is completely uncorrelated with respect to the efficiency of the process under study \textellipsis ''.

For a third example, Hollitt \citep[Appendix A]{Ho19} derives the uncertainty in ``purity'' independently of either Ullrich and Xu or Casadei and, unlike those two studies, without recourse to Bayes's rule, so without the need to postulate a prior probability. This purity appears similar to the others' efficiency. Hollitt obtains the same expression for uncertainty as Ullrich and Xu\textemdash their eq. (19)\textemdash describing the result of eq. \eqref{unc} above as ``the Frequentist prediction'', which is fair: the arguments in \S\ref{sec:IndepPF} are explicitly frequentist in nature (see e.g. \citep[\S5.1.1]{Bu21}), as is appropriate for the analytical situation herein and for the intended use of the resulting uncertainties.

Adoption of a frequentist approach is supported by a recent study of ROC-curve uncertainties in the context of machine learning \citep{Ur23}; in their \S9.1, the authors write that the ''baseline frequentist approach \ldots\enspace has correct [statistical] coverage by construction.''

\subsection{Aside \textemdash \ The Ends of a ROC Curve}
\label{sec:Aside} \noindent
Inaccuracies in the argument of \S\ref{sec:Effunc} enter not through use of eq. \eqref{uncProp} but rather eq. \eqref{varUncs}, which is the Gaussian approximation to a Poisson distribution, applying when $P$ and $F$ are large.

In the ROC curve shown in Fig. \ref{fig:ROC1}, which was constructed from the data in Fig. \ref{fig:two_histos}, the point closest to $(0,0)$ has $P=1199$ for the signal and $P=75$ for the background. Both are large enough for the Gaussian approximation to be sufficient; the lower limit is commonly cited to be about 10 (e.g. \citep[\S9.2.3]{Ur23}.  For the point closest to $(1,1)$, values of $F$ are many times larger.

The question of how to deal with very small values of $P$ and $F$ is not problematic for ROC-curve construction because the points $(0,0)$ and $(1,1)$ constitute opposite ends of every ROC curve, so they could, for example, be added by hand. From the point of view of comparing ROC curves, these points carry no information, precisely because they lie on all ROC curves. 

\section{Conclusion}
\label{sec:Conclusion} \noindent
Motivated by a desire to determine uncertainties on ROC (receiver operating characteristic) curves constructed from counting data, the calculation of uncertainty in efficiency is revisited; for the two coordinates of a point on a ROC curve are the efficiencies for signal and non-signal data. The derivation herein makes no assumption about the statistical nature of efficiency, nor does it require that a Bayesian prior be postulated. Rather, it relies on the conventional prescription for propagating uncertainties through a computation, which is no more than an adaptation of the standard expansion of a total derivative in terms of partial derivatives.

The expression obtained, eq. \eqref{unc}, has been derived before, likely many times, but some previous treatments (e.g. \citep{He09, Ul07, Ca12, Ho19}) criticise the result deprecatingly. This note argues to the contrary: that the result is correct for the type of efficiency used in the construction of a ROC curve from counting data. The authors cited use the term ``efficiency'' in other contexts; none mentions ROC curves. It seems that their efficiencies have a different statistical nature from those involved in ROC-curve construction. It is suggested that in this difference lies the source of the contention.

The authors of \citep{He09, Ul07, Ca12} express concern over the accuracy of eq. \eqref{unc} when the number of events either passing or failing the test is small. These conditions correspond to the two ends of a ROC curve. Uncertainties\textemdash even values\textemdash near these locations are unimportant from the point of view of distinguishing one ROC curve from another because all ROC curves have the same end-points.

\section*{Acknowledgment} \noindent
 This research was supported in part by the Australian Government through the Australian Research Council Centre of Excellence for Dark Matter Particle Physics (CDM, CE200100008).

\end{document}